\documentclass[12pt]{article}

\usepackage{appendix}
\usepackage{epsfig}
\usepackage{capt-of}
\usepackage{amsmath}

\usepackage[activeacute,english]{babel}
\usepackage{latexsym}

\newcommand{\bn}{\begin{eqnarray}}
\newcommand{\en}{\end{eqnarray}}
\newcommand{\be}{\begin{equation}}
\newcommand{\ee}{\end{equation}}

\newcommand{\bea}{\begin{eqnarray}}
\newcommand{\eea}{\end{eqnarray}}

\newcommand{\lsim}{\mbox{\raisebox{-.9ex}{~$\stackrel{\mbox{$<$}}{\sim}$~}}}
\newcommand{\gsim}{\mbox{\raisebox{-.9ex}{~$\stackrel{\mbox{$>$}}{\sim}$~}}}

\begin{document}

\title{\begin{flushright}
\normalsize PI/UAN-2008-314FT
\end{flushright}
\vspace{5mm}
{\bf On the Issue of the $\zeta$ Series Convergence and Loop Corrections in the Generation of Observable Primordial Non-Gaussianity in Slow-Roll Inflation. Part II: the Trispectrum}}



\author{
\hspace{-8mm}\textbf{
Yeinzon Rodr\'{\i}guez$^{1,2,}$\thanks{e-mail:
\texttt{yeinzon.rodriguez@uan.edu.co}}, C\'esar A. Valenzuela-Toledo$^{2,}$\thanks{e-mail: \texttt{cavalto@ciencias.uis.edu.co} }} \\ \\
\hspace{-12mm}\textit{$^1$Centro de Investigaciones, Universidad Antonio Nari\~no,}\\
\hspace{-12mm}\textit{Cra 3 Este \# 47A-15, Bogot\'a D.C., Colombia}\\
\hspace{-12mm}\textit{and} \\
\hspace{-12mm}\textit{$^2$Escuela de F\'{\i}sica, Universidad Industrial de Santander,}  \\
\hspace{-12mm}\textit{Ciudad Universitaria, Bucaramanga, Colombia} \\
}

\maketitle

\begin{abstract}

\noindent
We calculate the trispectrum $T_\zeta$ 
of the primordial curvature perturbation $\zeta$, generated during a {\it slow-roll} inflationary epoch by considering a two-field quadratic model 
of inflation with {\it canonical} kinetic terms. We consider loop contributions as well as tree-level terms, and show that it is possible to attain very high, {\it including observable}, values for the level of non-gaussianity  $\tau_{NL}$ if $T_\zeta$ is dominated by the one-loop contribution. 
Special attention is paid to the claim in JCAP {\bf 0902}, 017 (2009) [arXiv:0812.0807 [astro-ph]] that, in the 
model studied in this paper and for the specific inflationary trajectory we choose, the quantum fluctuations of the fields overwhelm the classical evolution. We argue that such a claim actually does not apply to our 
model, although more research is needed in order to understand the role of quantum diffusion. We also consider the probability that an observer in an ensemble of realizations of the density field sees a non-gaussian distribution. 
In that respect, we show that the probability associated to the chosen inflationary trajectory is non-negligible.
Finally, the levels of non-gaussianity $f_{NL}$ and $\tau_{NL}$ in the bispectrum $B_\zeta$ 
and trispectrum $T_\zeta$ 
of $\zeta$, respectively, are also studied for the case in which $\zeta$ is not generated during inflation.
\end{abstract}

\section{Introduction}

The primordial curvature perturbation $\zeta$ \cite{liddle,weinberg3,mukhanov,dodelson}, and its $\delta N$ expansion\footnote{By ``$\delta N$ expansion'', we mean approximating $\delta N$ by a power series expansion in the initial conditions. By ``$\delta N$ formula'', we mean the statement that to lowest order in spatial gradients $\zeta \equiv \delta N$. These conventions will be used throughout the text.} \cite{starobinsky,sasaki2,sata,lyth4,lyth2}, was the subject of study in a previous companion paper \cite{cogollo}. We were interested in how well the convergence of the $\zeta$ series was understood and if the traditional arguments to cut out the $\zeta$ series at second order \cite{lyth2,zaballa}, keeping only the tree-level terms to study the statistical descriptors of $\zeta$ \cite{alabidi1,vernizzi,battefeld,yokoyama1,yokoyama2,yokoyama3,seery3,byrnes2,byrnes3}, were reliable\footnote{In this paper, we follow the terminology of Ref. \cite{byrnes1} to identify the tree-level terms and the loop contributions in a diagrammatic approach. The associated diagrams are called {\it Feynman-like diagrams}.}. We argued that a previous study of the $\zeta$ series convergence, the viability of a perturbative regime, and the relative weight of the loop contributions against the tree-level terms were completely necessary and in some cases surprising. For instance, the levels of non-gaussianity $f_{NL}$ and $\tau_{NL}$ in the bispectrum $B_\zeta$ 
and trispectrum $T_\zeta$ 
of $\zeta$, respectively, for slow-roll inflationary models with canonical kinetic terms \cite{liddle,lyth5,lyth6}, are usually thought to be of order $\mathcal{O} (\epsilon_i,\eta_i)$ \cite{vernizzi,battefeld,yokoyama1}\footnote{See, however, Refs. \cite{alabidi1,byrnes3}.} and $\mathcal{O} (r)$ \cite{seery3,ssv}\footnote{See, however, Refs. \cite{slri,bch2}.}, respectively, where $\epsilon_i$ and $\eta_i$ are the slow-roll parameters with $\epsilon_i,|\eta_i| \ll 1$ \cite{lyth5}, 
and $r$ is the tensor to scalar ratio \cite{lyth6} with $r < 0.22$ at the $95 \%$ confidence level \cite{komatsu1}. However, in order to reach such a conclusion, 
generic models were used where the loop contributions are comparatively suppressed and, therefore, the truncated $\delta N$ expansion may be used. Of course exceptions may occur, and in those cases it is crucial to check up to what order the truncated $\delta N$ expansion may be used, and which loop contributions are larger than the tree-level terms.  In any of these cases, general models or exceptions, the question regarding the representation of $\zeta$ by the $\delta N$ expansion is a matter to discuss.
Refs. \cite{alabidi1,byrnes3} show that large, {\it and observable}, non-gaussianity in $B_\zeta$ is indeed possible for certain classes of {\it slow-roll} models with {\it canonical} kinetic terms and special trajectories in field space, relying only on the tree-level terms.  Ref. \cite{bch2} does the same for $B_\zeta$ and $T_\zeta$ but this time arguing that the loop corrections are always suppressed against the tree-level terms if the quantum fluctuations of the scalar fields do not overwhelm the classical evolution. Nonetheless, although the resultant phenomenology from papers in Refs. \cite{alabidi1,byrnes3,bch2} is very interesting, the classicality argument used in Ref. \cite{bch2} 
is very conservatively stated leading to conclusions that are too strong,
as we will argue later in this paper. 
More research remains to be done to understand the role of the quantum diffusion and, since this is beyond the scope of the present paper, we will leave the discussion for a
future research project. 
We addressed the $\zeta$ series convergence and the existence of a perturbative regime in the referred companion paper \cite{cogollo}, showing how important the requirements to guarantee those conditions are. Moreover, we showed that for a subclass of small-field {\it slow-roll} inflationary models with {\it canonical} kinetic terms, the one-loop correction to $B_\zeta$ might be much larger than the tree-level terms, giving as a result large, {\it and observable}, non-gaussianity parameterised by $f_{NL}$.  The present paper extends the analysis presented in Ref. \cite{cogollo} to $T_\zeta$ showing, for the first time, that {\it large and observable} non-gaussianity parameterised by $\tau_{NL}$ is possible in {\it slow-roll} inflationary models with {\it canonical} kinetic terms due to loop effects, in total contrast with the usual belief based on the results of Refs. \cite{seery3,ssv}. In order to properly identify the non-gaussianity levels found in Ref. \cite{cogollo} and in the present paper with those that are constrained by observation, we comment on the probability that an observer in an ensemble of realizations of the density field in our scenario sees a non-gaussian distribution. As we will show, such a probability is
non-negligible for the concave downward potential, making indeed the observation of the non-gaussianity studied in this paper quite possible. 


The layout of the paper is the following:  
in Section \ref{model}, we describe the 
slow-roll inflationary model that exhibits large levels of non-gaussianity when loop
corrections are considered.  In Section \ref{class}, we study the impact of the quantum
fluctuations of the scalar fields on their classical evolution.  As a result we argue how the
loop suppression proof given in Ref. \cite{bch2} does 
not apply to our model. Section \ref{prob} studies the probability of realizing the scenario
proposed in this paper for a typical observer.  Section \ref{constraints} is devoted to the
reduction of the available parameter window, for the most relevant case, by taking into
account some restrictions of
methodological and physical nature. The levels of non-gaussianity $f_{NL}$ in the bispectrum
$B_\zeta$ and $\tau_{NL}$ in the
trispectrum $T_\zeta$ are calculated in Section \ref{endcal} for models where $\zeta$ is, or
is not, generated during inflation; a comparison with the current literature and the results
found in the companion paper for $f_{NL}$ \cite{cogollo} is made.  
Finally, Section \ref{concl} presents the
conclusions.  The basic definitions employed in this paper and their relation with
observation are reviewed in Section 2 of Ref. \cite{cogollo}. Complementary cases to the ones
studied in Sections \ref{constraints} and \ref{endcal} are quickly developed in Appendix
\ref{app}.

\section{A quadratic two-field slow-roll model of inflation} \label{model}
We will study in this paper the inflationary potential given by
\begin{equation}
V = V_0\left(1+\frac{1}{2}\eta_\phi \frac{\phi^2}{m_P^2} +\frac{1}{2}\eta_\sigma \frac{\sigma^2}{m_P^2}\right) \,, \label{pot}
\end{equation}
where $\phi$ and $\sigma$ are the inflaton fields and $m_P$ is the reduced Planck mass.
By assuming that the first term in Eq. (\ref{pot}) dominates, $\eta_\phi$ and $\eta_\sigma$ become the usual $\eta$ slow-roll parameters associated with the fields $\phi$ and $\sigma$.

We have chosen the $\sigma=0$ trajectory, 
since this case is the easiest to work from the point of view of the analytical calculations and because it gives the most interesting results. In addition, such a trajectory for the potential in Eq. (\ref{pot}) reproduces for some number of e-folds (for $\eta_\phi,\eta_\sigma > 0$) the hybrid inflation scenario \cite{linde2}, where $\phi$ is the inflaton and $\sigma$ is the waterfall field.  We will analyze in Section \ref{prob} the probability for an observer to live in a region where $\sigma=0$ for the concave downward potential. Non-gaussianity in the bispectrum $B_\zeta$ of $\zeta$ for this kind of model has been studied in Refs. \cite{lyth2,cogollo,zaballa,alabidi1,byrnes3,bch2,enqvist1,vaihkonen,lyth3}; in particular, Ref. \cite{cogollo} shows that the one-loop correction dominates over the tree-level terms if $\eta_\phi,\eta_\sigma < 0$ and $|\eta_\sigma| > |\eta_\phi|$, generating in this way large values for $f_{NL}$ even if $\zeta$ {\it is} generated during inflation. Refs. \cite{alabidi1,byrnes3}, in contrast, work only at tree level with the same potential as Eq. (\ref{pot}) but relaxing the $\sigma = 0$ condition, finding that large values for $f_{NL}$ are possible for a small set of initial conditions. Ref. \cite{bch2} improves the analysis in Ref. \cite{byrnes3}, this time taking into account also the trispectrum $T_\zeta$ of $\zeta$ and the role of the loop corrections.  According to that reference, large values for $\tau_{NL}$ are also possible for a small set of initial conditions if the tree-level terms dominate over the loop corrections.  Moreover, it is claimed that loop corrections for this model are always suppressed against the tree-level terms if the quantum fluctuations of the fields are subdominant against their classical evolution.  The opposite case seems to happen for some narrow range of initial conditions including $\sigma_\star = 0$, which is the case studied in this paper. 
As we will argue in Section \ref{class}, the classicality condition in Ref. \cite{bch2} is expressed in a very conservative way, leading to too strong and nongeneral conclusions.  Dominance of loop corrections is then safe from the classical vs quantum condition, allowing the interesting large levels of non-gaussianity discussed in Ref. \cite{cogollo} and in the present paper.

Since we are considering a slow-roll regime, the evolution of the fields for the 
inflationary model given by the potential in Eq. (\ref{pot}) is
\bea
\phi(N) &=& \phi_\star \exp(-N\eta_\phi)\,, \label{srp} \\
\sigma(N)&=&\sigma_\star \exp(-N\eta_\sigma)\,, \label{srs}
\eea
where $\phi_\star$ and $\sigma_\star$ are the initial field values at the time when the relevant cosmological scales leave the horizon. Thus, in the framework of the $\delta N$ formalism \cite{starobinsky,sasaki2,sata,lyth4,lyth2}, the potential in Eq. (\ref{pot}) leads to the following derivatives of $N$ with respect to $\phi_\star$ and $\sigma_\star$ (evaluated in $\sigma_\star = 0$):
\bea
N_\phi &=& \frac{1}{\eta_\phi \phi_\star} \,, \hspace{5mm}  N_\sigma = 0 \,, \label{1d} \\
N_{\phi \phi} &=& -\frac{1}{\eta_\phi \phi_\star^2} \,, \hspace{5mm} N_{\phi \sigma} = 0 \,, \hspace{5mm} N_{\sigma \sigma} = \frac{\eta_\sigma}{\eta_\phi^2 \phi_\star^2} \exp[2N (\eta_\phi - \eta_\sigma)] \,. \label{2d}
\eea
The above derivatives are then used to calculate the leading terms to the spectrum, bispectrum, and trispectrum of the primordial curvature perturbation $\zeta$ including the tree-level and one-loop contributions:
\bea
{\mathcal P}_\zeta^{tree}&=&\frac{1}{\eta_\phi^2\phi_\star^2}\left(\frac{H_\star}{2 \pi}\right)^2 \,,\label{pt}\\
{\mathcal P}_\zeta^{ 1-loop}&=&\frac{\eta_\sigma^2}{\eta_\phi^4\phi_\star^4}\exp[4N(\eta_\phi-\eta_\sigma)] \left(\frac{H_\star}{2 \pi}\right)^4\ln(kL) \,,\label{pl}\\
B_\zeta^{tree} &=& -\frac{1}{\eta_\phi^3 \phi_\star^4} \left(\frac{H_\star}{2\pi}\right)^4 4\pi^4 \left(\frac{\sum_i k_i^3}{\prod_i k_i^3}\right) \,, \label{bt} \\
B_\zeta^{1-loop} &=& \frac{\eta_\sigma^3}{\eta_\phi^6 \phi_\star^6} \exp[6N(\eta_\phi - \eta_\sigma)] \left(\frac{H_\star}{2\pi}\right)^6 \ln(kL) 4\pi^4 \left(\frac{\sum_i k_i^3}{\prod_i k_i^3}\right) \,, \label{bl}\\
T_\zeta^{tree}&=& \frac{1}{\eta_\phi^4\phi_\star^6}\left(\frac{H_\star}{2 \pi}\right)^6\left[\frac{2\pi^2}{k_2^3}\frac{2\pi^2}{k_4^3}\frac{2\pi^2}{|{\bf k}_3+{\bf k}_4|^3} + 11 \ {\rm permutations}\right] \,,\label{tt}\\
T_\zeta^{ 1-loop}&=&\frac{\eta_\sigma^4}{\eta_\phi^8\phi_\star^8}\exp[8N(\eta_\phi-\eta_\sigma)] \left(\frac{H_\star}{2 \pi}\right)^8\ln(kL) \ 4 \Big[ \frac{2\pi^2}{k_2^3}\frac{2\pi^2}{k_4^3}\frac{2\pi^2}{|{\bf k}_3+{\bf k}_4|^3} + \nonumber \\
&&+11 \ {\rm permutations} \Big] \,,\label{tl}
\eea
where $L$ is the infrared cutoff chosen so that the quantities are calculated in a minimal
box \cite{lyth1,bernardeu4}. Except when considering low CMB multipoles, the box size should
be set at $L \sim H_0$ \cite{leblond,kohri}\footnote{$H_0$ is the Hubble parameter today.},
giving $\ln(kL) \sim \mathcal{O} (1)$ for relevant cosmological scales.

The important factor in the loop corrections is the exponential.  This exponential function is directly related to the quadratic form of the potential with a leading constant term.  It will give a large contribution if $\eta_\phi > \eta_\sigma$.  In Ref. \cite{cogollo}, we chose the concave downward potential in order to satisfy the spectral tilt constraint, which makes $\eta_\phi < 0$, 
while keeping $|\eta_\sigma| > |\eta_\phi|$.  In this paper, we will consider 
the same case.

\section{Classicality} \label{class}

Ref. \cite{bch2} argues in Appendices A and B how, by imposing the requirement that the quantum fluctuations of the fields around their background values do not overwhelm the respective classical evolutions, the loop corrections to $P_\zeta$, $B_\zeta$, and $T_\zeta$ are suppressed against the tree-level terms. The proof relies on the fact that, if the classicality condition is satisfied, the second-order terms in the $\delta N$ expansion \cite{lyth4,lyth2}
\begin{eqnarray}
\zeta(t,\textbf{x})&=&\sum_{i}\textit{N}_{i}(t)\delta\phi_{i}(t_\star,{\bf x}) - \sum_{i}\textit{N}_{i}(t) \langle \delta\phi_{i}(t_\star,{\bf x}) \rangle + \nonumber \\
&& + \frac{1}{2}\sum_{ij}\textit{N}_{ij}(t)\delta\phi_{i}(t_\star,{\bf x})\delta\phi_{j}(t_\star,{\bf x}) - \frac{1}{2}\sum_{ij}\textit{N}_{ij}(t) \langle \delta\phi_{i}(t_\star,{\bf x})\delta\phi_{j}(t_\star,{\bf x}) \rangle + \nonumber \\
&& + \frac{1}{3!}\sum_{ijk}\textit{N}_{ijk}(t)\delta\phi_{i}(t_\star,{\bf x})\delta\phi_{j}(t_\star,{\bf x})
\delta\phi_{k}(t_\star,{\bf x}) - \frac{1}{3!}\sum_{ijk}\textit{N}_{ijk}(t) \langle \delta\phi_{i}(t_\star,{\bf x})\delta\phi_{j}(t_\star,{\bf x})
\delta\phi_{k}(t_\star,{\bf x}) \rangle + \nonumber \\
&& + ...\;, \label{Nexp}
\end{eqnarray}
are subleading against the first-order terms\footnote{In the previous expression, $t_\star$ denotes the time when the relevant cosmological scales exit the horizon.}. This in turn implies
\begin{eqnarray}
\frac{P_\zeta^{1-loop}}{P_\zeta^{tree}} \ll 1 \,, \\
\frac{B_\zeta^{1-loop}}{B_\zeta^{tree}} \ll 1 \,, \\ 
\frac{T_\zeta^{1-loop}}{T_\zeta^{tree}} \ll 1 \,,
\end{eqnarray}
as explicitly stated in Eqs. (A.16-A.19) of Ref. \cite{bch2}.  In addition, under the same assumptions, higher order corrections in the spectral functions $P_\zeta$, $B_\zeta$, and $T_\zeta$ are always subleading against the one-loop corrections and, therefore, subleading against the tree-level terms. This conclusion is obtained if the $\delta N$ expansion may be truncated at fourth order. However, what is the classicality condition employed in Ref. \cite{bch2}?

Assuming slow-roll evolution for each field $\phi_i$, which is valid only if the quantum fluctuation $\delta \phi_i$ is by far smaller than the classical evolution $\Delta \phi_i$, the classical change in the $\phi_i$ field during a Hubble time around horizon exit is
\begin{equation}
\Delta \phi_i (t_\star) \approx -\frac{V_i (\phi)}{3H_\ast^2 \sqrt{6}} \,, 
\end{equation}
where $V_i$ denotes the derivative of the potential with respect to the $i$-th field. Comparing the latter expression with the quantum fluctuation
\begin{equation}
\delta \phi_i (t_\star) \approx \frac{H_\ast}{2\pi} \,, 
\end{equation}
and requiring that $\Delta \phi_i$ is much larger than $\delta \phi_i$, we get
\begin{equation}
|\dot{\phi_i}|_\star \gg \sqrt{\frac{3}{2\pi^2}} H_\ast^2 \,. \label{classcondhy}
\end{equation}
For our quadratic two-field slow-roll model of inflation, where the slow-roll evolution is given by Eqs. (\ref{srp})-(\ref{srs}), the previous expression translates into
\begin{equation}
|\phi_i|_\star \gg \sqrt{\frac{3}{2\pi^2}} \left|\frac{H_\ast}{\eta_i}\right| \,, 
\end{equation}
which is the one given in Eq. (A.1) of Ref. \cite{bch2}. Such a condition is equivalent to
\begin{equation}
\left|\frac{\delta \phi_i}{\phi_i}\right|_\star \ll \left|\frac{\eta_i}{\sqrt{6}}\right| \,, \label{classcond} 
\end{equation}
which is the one given in Eq. (A.2) of Ref. \cite{bch2}. Thus, under this condition, the trajectory $\sigma = 0$ studied in this paper and its companion \cite{cogollo} seems not to be well described by the slow-roll approximation and, therefore, the obtained results based in the $\delta N$ formalism would not be reliable.

This classicality argument given in Ref. \cite{bch2} is too conservatively stated. To see why it is like that,
we may reason in the following way for general inflationary models: 
for any point along the background classical trajectory in field space it is possible to rotate the
field axes so that, instantaneously, there is an inflaton (or `adiabatic') field that points
along the trajectory and some light `entropy' fields which point in orthogonal directions
\cite{gordon}. The quantum fluctuations for the entropy fields are nonvanishing, but the
classical evolution for each of these fields is zero.  Since the condition in
Eq. (\ref{classcondhy}) is not formulated in any particular field parameterisation, we may
argue that for any multifield inflationary model the application of this condition would
lead to a background inflationary trajectory dominated by the quantum evolution. Thus,
slow-roll conditions would always be impossible to apply. 
A more general classicality condition should still be the one in Eq. (\ref{classcondhy}) but
only applied to the adiabatic field and not to the entropy fields. In that respect, the
$\sigma = 0$ trajectory studied in this paper and its companion \cite{cogollo} is safe from
large quantum fluctuations, since the classicality condition in Eq. (\ref{classcondhy})
applied only to the $\phi$ field is extremely well satisfied as long as $P_\zeta \ll 1$,
which is actually the case for single field slow-roll inflation \cite{thesis}. Indeed, a
much better way of stating the classicality condition is the
following:\footnote{We acknowledge Misao Sasaki for pointing out to us this idea.} if the
inflationary trajectory must be dominated by the classical motion of the fields, then the
perturbation in the amount of inflation, due to the quantum fluctuations of the fields,
must be negligible:
\begin{equation}
\delta N \ll 1 \,. 
\end{equation}
By virtue of the $\delta N$ formalism, this expression is simply satisfied if the free
parameters of the inflationary model under consideration are chosen so that the COBE
normalisation ($\mathcal{P}_\zeta^{1/2} \approx 5 \times 10^{-5}$ \cite{bunn}) is satisfied,
which is always the case. Nevertheless, we understand that the role of quantum diffusion is
of great importance (see, for instance, Refs. \cite{garcia-bellido,randall}), and a dedicated study of this issue is
left for a future research project.

\section{Probability} \label{prob}

The main purpose of this paper is to identify regions in the parameter space with high levels of primordial non-gaussianity. Then, we proceed to compare the obtained non-gaussianity with observation. In order to do the latter, we first need to realize what the probability is for a typical observer to live in a universe where the inflationary trajectory is the one studied in this paper: $\sigma = 0$.  This is particularly relevant for the concave downward potential where the background trajectory $\sigma = 0$ is unstable.

In the context of quantum cosmology, the probability of quantum creation of a closed universe is proportional to \cite{lindeprob,vilenkin1,vilenkin2,vilenkin3}
\begin{equation}
P \sim \exp \left(-\frac{24\pi^2 m_P^4}{V}\right) \,, 
\end{equation}
which means that the Universe can be created if $V$ is not too much smaller than the Planck density.  Thus, for our concave downward potential, having chosen the field contributions to $V$ in Eq. (\ref{pot}) to be negligible is good, because it increases the probability. In addition, within a set of initial conditions for $\phi$, the most probable initial condition for $\sigma$ is $\sigma = 0$.  The $\phi = 0$ trajectory is also highly probable but, since we are assuming $|\eta_\sigma| > |\eta_\phi|$, the $\sigma = 0$ trajectory is more probable. This of course implies that the levels of non-gaussianity obtained in this paper may be observable.



\section{Reducing the available parameter window} \label{constraints}

\begin{figure}
\begin{center}
\begin{tabular}{cccc}
\includegraphics[width=7cm,height=2cm]{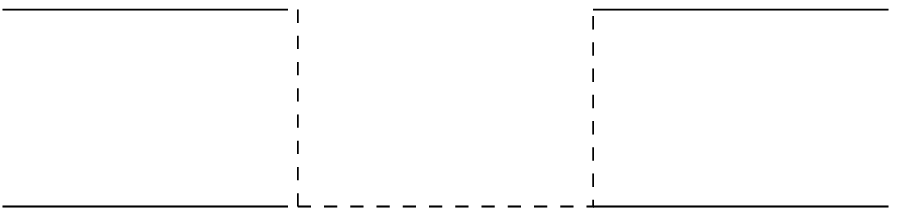} & & & \includegraphics[width=7cm,height=2cm]{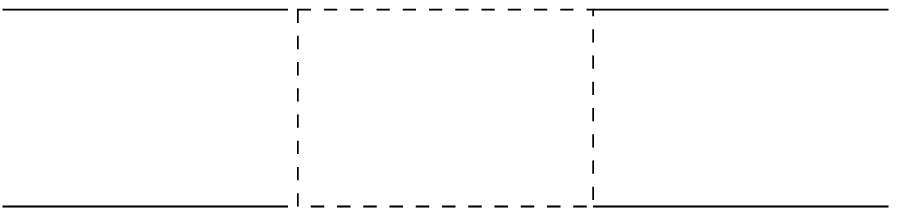} \\
(a) & & & (b)
\end{tabular}
\end{center}
\caption{(a). Tree-level Feynman-like diagram for $T_\zeta$. (b). One-loop Feynman-like diagram for $T_\zeta$. The internal dashed lines correspond to two-point correlators of field perturbations.} \label{tf}
\end{figure}

The analysis of the observed spectral index and the $\zeta$ series convergence are given in Subsection 5.3 and Section 7 of Ref. \cite{cogollo}, respectively.  Regarding the existence of a perturbative regime, we have to add to the discussion in Section 7 of Ref. \cite{cogollo} that, by cutting out the series in Eq. (98) of Ref. \cite{cogollo} at second order, just one Feynman-like diagram per spectral function of $\zeta$ is necessary to study the loop corrections to these spectral functions\footnote{This is assuming that the Feynman-like diagrams containing $n$-point correlators of the field perturbations with $n \geq 3$ are subdominant against the diagrams containing only two-point correlators.  For the trispectrum this does not happen when the tree-level terms dominate over the loop corrections \cite{seery3,ssv}. However, for the cases considered in this paper, when the loop corrections in the trispectrum dominate over the tree-level terms generating in turn large values for $\tau_{NL}$, the diagrams containing $n$-point correlators of the field perturbations with $n \geq 3$ are expected to be subdominat because of their dependence on the slow-roll parameters \cite{jarnhus}.}. That is why in the companion paper \cite{cogollo} there was just one leading diagram for the one-loop correction to $P_\zeta$ (Fig. A.2(a) of Ref. \cite{cogollo}), as well as one leading diagram for the one-loop correction to $B_\zeta$ (Fig. A.4(a) of Ref. \cite{cogollo}). The same argument can be used to justify not to have considered the dressed vertices diagrams \cite{byrnes1} in Ref. \cite{cogollo}, since these diagrams imply at least considering third-order terms in the $\zeta$ series expansion. When applied to $T_\zeta$, this analysis shows that the only diagrams to consider are the one in Fig. \ref{tf}a for the tree-level terms, and the one in Fig. \ref{tf}b for the loop corrections. Such diagrams lead to the expressions in Eqs. (\ref{tt}) and (\ref{tl}) for $T_\zeta^{tree}$ and $T_\zeta^{ 1-loop}$.
In the following, we will give the relevant information for $T_\zeta$ when $\zeta$ {\it is} generated during inflation, and for both $B_\zeta$ and $T_\zeta$ when $\zeta$ {\it is not} generated during inflation. 

\subsection{Tree-level or loop dominance}
The exponential factors in Eqs. (\ref{pl}) and (\ref{tl}) open up the possibility that the
loop corrections dominate over  $\mathcal{P}_\zeta$ and/or $T_\zeta$. There are three
posibilities: 1. both $T_\zeta$ and $\mathcal{P}_\zeta$ are dominated by the one-loop corrections,
2. $T_\zeta$ is dominated by the one-loop correction and $\mathcal{P}_\zeta$ is dominated by
the tree-level term, and 3. both $T_\zeta$ and $\mathcal{P}_\zeta$ are dominated by the tree-level
terms.  However, we will concentrate on the most relevant case (second possibility), leaving
the discussion regarding the other possibilities for Appendix \ref{app}.

\subsubsection{$T_\zeta$ dominated by the one-loop correction and $\mathcal{P}_\zeta$ dominated by the tree-level term} \label{t1lptresub}
Comparing Eqs. (\ref{pt}) with (\ref{pl}) and Eqs. (\ref{tt}) with (\ref{tl}), we require in this case that
\bea
\frac{\eta_\sigma^2}{\eta_\phi^2} \exp[4N(\eta_\phi-\eta_\sigma)] &\ll& \frac{1}{\frac{1}{\phi_\star^2} \left(\frac{H_\star}{2\pi}\right)^2} \,, \label{t1lptre} \\
4\frac{\eta_\sigma^4}{\eta_\phi^4} \exp[8N(\eta_\phi-\eta_\sigma)] &\gg& \frac{1}{\frac{1}{\phi_\star^2} \left(\frac{H_\star}{2\pi}\right)^2} \,. \label{t1lptre2}
\eea
Employing the definition for the tensor to scalar ratio $r$ \cite{lyth6},
\be\label{rts}
r\equiv\frac{{\mathcal P}_T}{{\mathcal P}_\zeta^{\rm obs}}=\frac{\frac{8}{m^2_P}\left(\frac{H_\star}{2 \pi}\right)^2}{{\mathcal P}_\zeta^{\rm obs}}\,,
\ee
where
${\mathcal P}^{1/2}_T$ is the amplitude of the spectrum for primordial gravitational waves and
${\mathcal P}_\zeta^{\rm obs}$ is the {\it observed} primordial curvature perturbation spectrum;
we may combine Eqs. (\ref{t1lptre}) and (\ref{t1lptre2}) as
\be
\frac{r \mathcal{P}_\zeta^{\rm obs}}{8} \frac{\eta_\sigma^2}{\eta_\phi^2} \exp[4N(\eta_\phi-\eta_\sigma)] \ll \left(\frac{\phi_\star}{m_P}\right)^2 \ll \frac{r \mathcal{P}_\zeta^{\rm obs}}{8} \frac{4\eta_\sigma^4}{\eta_\phi^4} \exp[8N(\eta_\phi-\eta_\sigma)] \,.
\label{intc}
\ee
From now on, we will name the parameter window described by Eq. (\ref{intc}) as the
intermediate $\phi_\star$ $T$-region\footnote{The $T$ in $T$-region is introduced in this
paper in order to differentiate explicitly between these regions and those found in the
companion paper \cite{cogollo} for $B_\zeta$.}, since the latter represents a region of
allowed values for $\phi_\star$ limited by both an upper and a lower bound.

\subsection{The normalisation of the spectrum}
Either $\zeta$ is or is not generated during inflation, we must satisfy the appropriate
spectrum normalisation condition. Four possibilities exist: 1. $\zeta$ is generated
during inflation and $\mathcal{P}_\zeta$ is dominated by the one-loop correction, 2. $\zeta$
is not generated during inflation and $\mathcal{P}_\zeta$ is dominated by the one-loop
correction, 3. $\zeta$ is generated during inflation and $\mathcal{P}_\zeta$ is dominated by
the tree-level term, and 4. $\zeta$ is not generated during inflation and $\mathcal{P}_\zeta$
is dominated by the tree-level term. However, it was shown in Ref. \cite{cogollo} that the 
first possibility is of no observational interest, since it is impossible to reproduce the
observed spectral index and its running. Among the other three possibilities, the ones we are
interested in the most are the third and fourth. We will study such possibilities right now,
leaving the other one for Appendix \ref{app}.

\subsubsection{$\zeta$ generated during inflation ($\mathcal{P}_\zeta = \mathcal{P}_\zeta^{\rm obs}$) and $\mathcal{P}_\zeta$ dominated by the tree-level term} \label{pnortrsub}
According to Eqs. (\ref{pt}) and (\ref{rts}) we have in this case
\bea
\mathcal{P}_\zeta^{tree} &=& \frac{1}{\eta_\phi^2 \phi_\star^2} \left(\frac{H_\star}{2\pi}\right)^2 \nonumber \\
&=& \frac{1}{\eta_\phi^2} \left(\frac{m_P}{\phi_\ast}\right)^2 \frac{r \mathcal{P}_\zeta}{8} \,, \label{pnortr}
\eea
which reduces to
\be
\left(\frac{\phi_\star}{m_P}\right)^2 = \frac{1}{\eta_\phi^2} \frac{r}{8} \,. \label{normt}
\ee
Notice that in such a situation, the value of the $\phi$ field when the relevant scales are exiting the horizon depends exclusively on the tensor to scalar ratio, once $\eta_\phi$ has been fixed by the spectral tilt constraint. 

\subsubsection{$\zeta$ not generated during inflation ($\mathcal{P}_\zeta \ll \mathcal{P}_\zeta^{\rm obs}$) and $\mathcal{P}_\zeta$ dominated by the tree-level term}
The previous subsubsection tells us that in this case the constraint to satisfy is
\be\label{normtd}
\left(\frac{\phi_\star}{m_P}\right)^2 \gg \frac{1}{\eta_\phi^2} \frac{r}{8} \,.
\ee

\subsection{End of inflation}
Because of the characteristics of the inflationary potential in Eq. (\ref{pot}), inflation is
eternal in this model.  However, Ref. \cite{naruko} introduced the multibrid inflation idea
of Refs. \cite{mb1,mb2} so that the potential in Eq. (\ref{pot}) is achieved during inflation
while a third field $\rho$ acting as a waterfall field is stabilized in $\rho = 0$. During
inflation, $\rho$ is heavy and it is trapped with a vacuum expectation value equal to zero, so
neglecting it during inflation is a good approximation. The end of inflation comes when the
effective mass of $\rho$ becomes negative, 
which is possible to obtain if $V_0$ in the potential of Eq. (\ref{pot}) is replaced by
\begin{equation}
V_0 = \frac{1}{2} G(\phi,\sigma) \rho^2 + \frac{\lambda}{4} \left(\rho^2 - \frac{\varSigma^2}{\lambda} \right)^2 \,, \label{endcoup}
\end{equation}
where 
\begin{equation}
G(\phi,\sigma) \equiv g_1\phi^2 + g_2\sigma^2 \,, \label{hypend}
\end{equation}
$\varSigma$ has some definite value, 
and $g_1$, $g_2$, and $\lambda$ are some coupling constants.
The end of inflation actually happens
on a hypersurface defined in general
by \cite{naruko,mb1}
\begin{equation}
G(\phi,\sigma) = \varSigma^2 \,. 
\end{equation}

In general the hypersurface in Eq. (\ref{hypend}), defined by the end of inflation condition,
is not a surface of uniform energy density.  Because the $\delta N$ formalism requires the
final slice to be of uniform energy density\footnote{See, for instance, Ref. \cite{cogollo}.},
we need to add a small correction term to the amount of expansion up to the surface where
$\rho$ is destabilised.  In addition, the end of inflation is inhomogeneous, which
generically leads to different predictions from those obtained during inflation for the
spectral functions \cite{lythend,salem,alabidiend}. In particular, large levels of
non-gaussianity may be obtained by tuning the free parameters of the model. Specifically,
by making $g_1/g_2 \ll 1$, large values for $f_{NL}$ and $\tau_{NL}$ are obtained due to the
end of inflation mechanism rather than due to the dynamics during slow-roll inflation
\cite{naruko,alabidiend,huang}.

Ref. \cite{bch2} chose instead the case $g_1^2/g_2^2 = \eta_\phi/\eta_\sigma$ such that the surface where $\rho$ is destabilised corresponds to a surface of uniform energy density\footnote{Ref. \cite{bch2} studies as well the case where $g_1^2 = g_2^2$, but we are not going to consider it here.}. In this case, all of the spectral functions are the same as those calculated in this paper and in Refs. \cite{cogollo,bch2}, which in turn are valid at the final hypersurface of uniform energy density during slow-roll inflation. Thus, we have a definite mechanism to end inflation which, nevertheless, leaves intact the non-gaussianity generated during inflation.

It is well known that the number of e-folds of expansion from the time the cosmological scales exit the horizon to the end of inflation is presumably around but less than 62 \cite{liddle,weinberg3,mukhanov,dodelson}.  The slow-roll evolution of the $\phi$ field in Eq. (\ref{srp}) tells us that such an amount of inflation is given by
\begin{equation}
N = -\frac{1}{\eta_\phi} \ln\left(\frac{\phi_{end}}{\phi_\star}\right) \lsim 62 \,, \label{end}
\end{equation}
where $\phi_{end}$ is the value of the $\phi$ field at the end of inflation. Such a value depends noticeably on the coupling constants in Eq. (\ref{endcoup}).  We will in this paper not concentrate on the allowed parameter window for $g_1$, $g_2$, and $\lambda$. Instead, we will give an upper bound on $\phi$ during inflation, for the $\eta_\phi < 0$ case, consistent with the potential in Eq. (\ref{pot}) and the end of inflation mechanism described above. 


Keeping in mind the results of Ref. \cite{armendariz2} which say that the ultraviolet cutoff in cosmological perturbation theory could be a few orders of magnitude bigger than $m_P$, we will tune the coupling constants in Eq. (\ref{endcoup}) so that inflation for $\eta_\phi < 0$ comes to an end when $|\eta_\phi|\phi^2/2m_P^2 \sim 10^{-2}$. This allows us to be on the safe side (avoiding large modifications to the potential coming from ultraviolet cutoff-suppressed nonrenormalisable terms, and keeping the potential dominated by the constant $V_0$ term). 
Coming back to Eq. (\ref{end}), we get then
\begin{equation}
N = \frac{1}{|\eta_\phi|} \ln\left[\left(\frac{2\times10^{-2}}{|\eta_\phi|}\right)^{1/2}\frac{m_P}{\phi_\star}\right] \lsim 62 \,, \label{amount}
\end{equation}
which leads to
\begin{equation}
\frac{\phi_\star}{m_P} \gsim \left(\frac{2\times10^{-2}}{|\eta_\phi|}\right)^{1/2}\exp(-62|\eta_\phi|) \,. \label{amountc}
\end{equation}

\section{$f_{NL}$ and $\tau_{NL}$} \label{endcal}
In this section, we will calculate the levels of non-gaussianity represented in the parameters
$f_{NL}$ and $\tau_{NL}$
defined as \cite{maldacena,boubekeur1}
\bea
\frac{6}{5} f_{NL} &=& \frac{B_\zeta}{4\pi^4 \frac{\sum_i k_i^3}{\prod_i k_i^3} \mathcal{P}_\zeta^2} \,, \\
\frac{1}{2} \tau_{NL} &=& \frac{T_\zeta}{8\pi^6 \left[\frac{1}{k_2^3 k_4^3 |{\bf k}_3 + {\bf k}_4|^3} + 23 \;\; {\rm permutations}\right] \mathcal{P}_\zeta^3} \,. \label{tau}
\eea

\subsection{$\zeta$ generated during inflation ($\mathcal{P}_\zeta = \mathcal{P}_\zeta^{\rm obs}$)}

\subsubsection{The intermediate $\phi_\star$ $T$-region}
The level of non-gaussianity $\tau_{NL}$ according to Eqs. (\ref{pt}), (\ref{tl}) and (\ref{tau}), is given by
\bea
\frac{1}{2} \tau_{NL} &=& \frac{T^{1-loop}_\zeta}{8\pi^6 \left[\frac{1}{k_2^3 k_4^3 |{\bf k}_3 + {\bf k}_4|^3} + 23 \;\; {\rm permutations}\right] (\mathcal{P}^{tree}_\zeta)^3} \nonumber \\
&=&\frac{2\eta_\sigma^4}{\eta_\phi^2\phi_\star^2}\exp[8N(|\eta_\sigma|-|\eta_\phi|)] \left(\frac{H_\star}{2\pi}\right)^2\ln(kL)\nonumber\\
&=&\frac{2 \eta_\sigma^4}{\eta_\phi^2}\exp[8N(|\eta_\sigma|-|\eta_\phi|)]\left(\frac{m_P}{\phi_\star}\right)^2\frac{r{\mathcal P}_\zeta}{8}\ln(kL)\nonumber\\
&=&2 \eta_\sigma^4\exp[8N(|\eta_\sigma|-|\eta_\phi|)]{\mathcal P}_\zeta\ln(kL)\nonumber\\
\Rightarrow\;\;\frac{1}{2}\tau_{NL}&\simeq& 4.91\times 10^{-9}|\eta_\sigma|^4\exp[400\ln(5.657\times10^{-2}r^{-1/2})(|\eta_\sigma|-0.020)]\,,
\eea
where in the last line we have used the expressions in Eqs. (\ref{normt}) and (\ref{amount}) of this paper, and in Eq. (78) of Ref. \cite{cogollo}.

\begin{figure*} [t]
\begin{center}
\includegraphics[width=15cm,height=10cm]{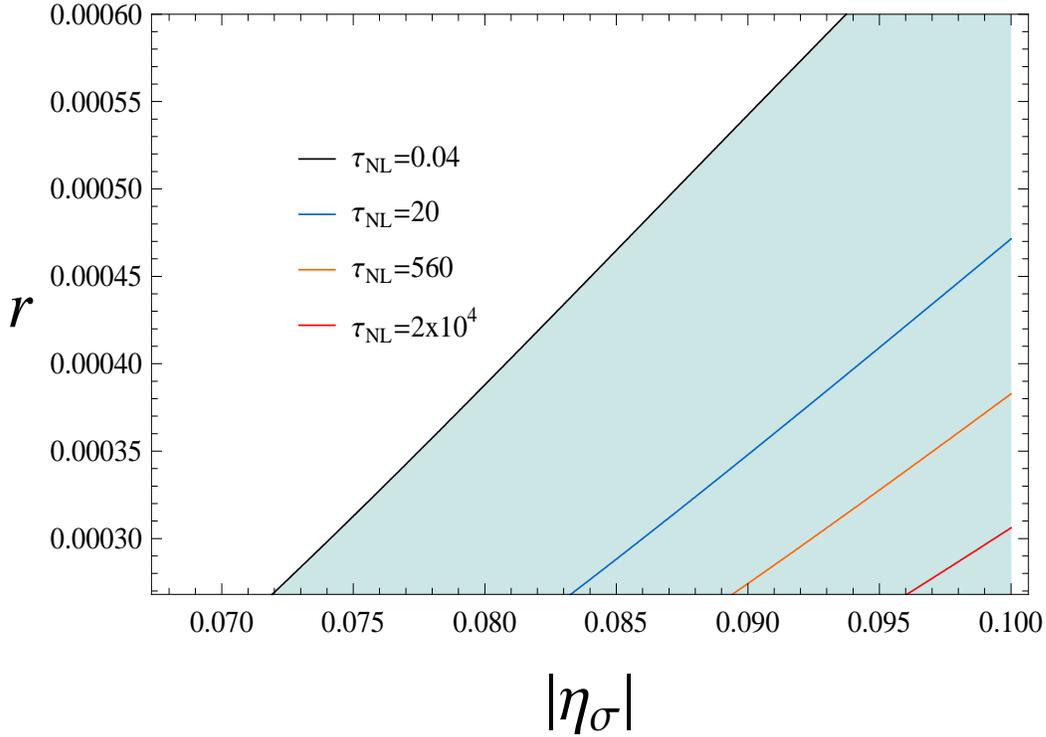}
\end{center}
\caption{Contours of $\tau_{NL}$ in the $r$ vs $|\eta_\sigma|$ plot. The intermediate (high)
$\phi_\star$ $T$-region corresponds to the shaded (white) region. The observationally
expected $2\sigma$ range of values, for WMAP, PLANCK, and even the 21 cm background
anisotropies, and for positive $\tau_{NL}$, $\tau_{NL} > 20$ are completely inside the
intermediate $\phi_\star$ $T$-region.  Notice that the boundary line between the high (see
Subsubsection \ref{hiphit}) and
the intermediate $\phi_\star$ $T$-regions matches almost exactly the $\tau_{NL} = 0.04$ line.}
\label{fig1}
\end{figure*}

Now, by implementing the spectral tilt constraint in Eq. (78) of Ref. \cite{cogollo} in the spectrum normalisation constraint in Eq. (\ref{normt}) and the amount of inflation constraint in Eq. (\ref{amountc}), we conclude that the tensor to scalar ratio $r$ is bounded from below  $r \gsim 2.680 \times 10^{-4}$. 

In the $r$ vs $|\eta_\sigma|$ plot in figure \ref{fig1}, we show lines of constant
$\tau_{NL}$ corresponding to the values $\tau_{NL} = 20, 560, 2 \times 10^4$.  We also show
the high (in white) (see Subsubsection \ref{hiphit}) and intermediate (shaded) $\phi_\star$ $T$-regions in agreement with the constraint in Eq. (\ref{intc}):
\bea
&&\frac{r \mathcal{P}_\zeta}{8} \frac{\eta_\sigma^2}{\eta_\phi^2} \exp[4N(|\eta_\sigma|-|\eta_\phi|)] \ll \left(\frac{\phi_\star}{m_P}\right)^2 \ll \frac{r \mathcal{P}_\zeta}{2} \frac{\eta_\sigma^4}{\eta_\phi^4} \exp[8N(|\eta_\sigma|-|\eta_\phi|)] \,,\nonumber\\
&\Rightarrow&4.070\times10^4\ll|\eta_\sigma|^4\exp[400\ln(5.657\times10^{-2}r^{-1/2})(|\eta_\sigma|-0.020)]\ll1.656\times10^{17} \,.
\eea
As is evident from the plot, the observationally expected $2\sigma$ range of values for WMAP, $|\tau_{NL}| \gsim 2 \times 10^4$ \cite{kogo}, PLANCK, $|\tau_{NL}| \gsim 560$ \cite{kogo}, and even the 21 cm background anisotropies, $|\tau_{NL}| \gsim 20$ \cite{cooray2}, and for positive $\tau_{NL}$, are completely inside the intermediate $\phi_\star$ $T$-region as required.  Higher values for $\tau_{NL}$, up to $\tau_{NL} = 1.7 \times 10^5$ are consistent within our framework for the intermediate $\phi_\star$ $T$-region. 

\begin{figure*} [t]
\begin{center}
\includegraphics[width=15cm,height=10cm]{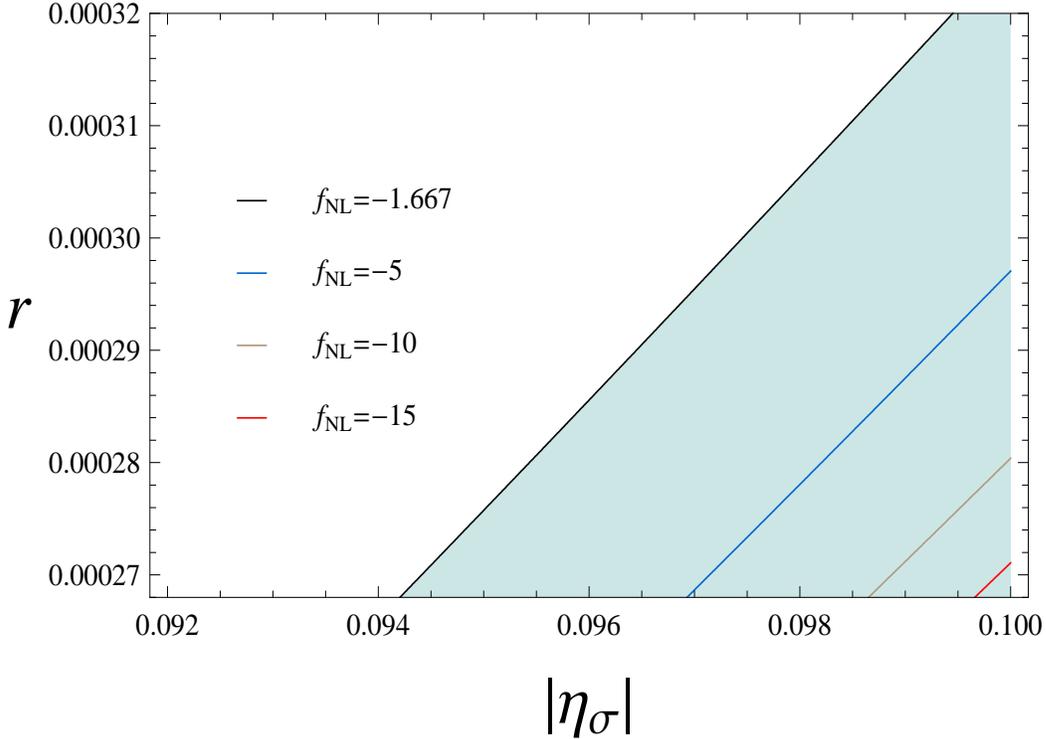}
\end{center}
\caption{Contours of $f_{NL}$ in the $r$ vs $|\eta_\sigma|$ plot. The intermediate (high) $\phi_\star$ region corresponds to the shaded (white) region. The WMAP (and also PLANCK) observationally allowed $2\sigma$ range of values for negative $f_{NL}$, $-9 < f_{NL}$, is completely inside the intermediate $\phi_\star$ region. Notice that the boundary line between the high and the intermediate $\phi_\star$ regions matches almost exactly the $f_{NL} = -1.667$ line. (This figure has been taken from Ref. \cite{cogollo}).}
\label{fig2}
\end{figure*}

\begin{figure*} [t]
\begin{center}
\includegraphics[width=15cm,height=10cm]{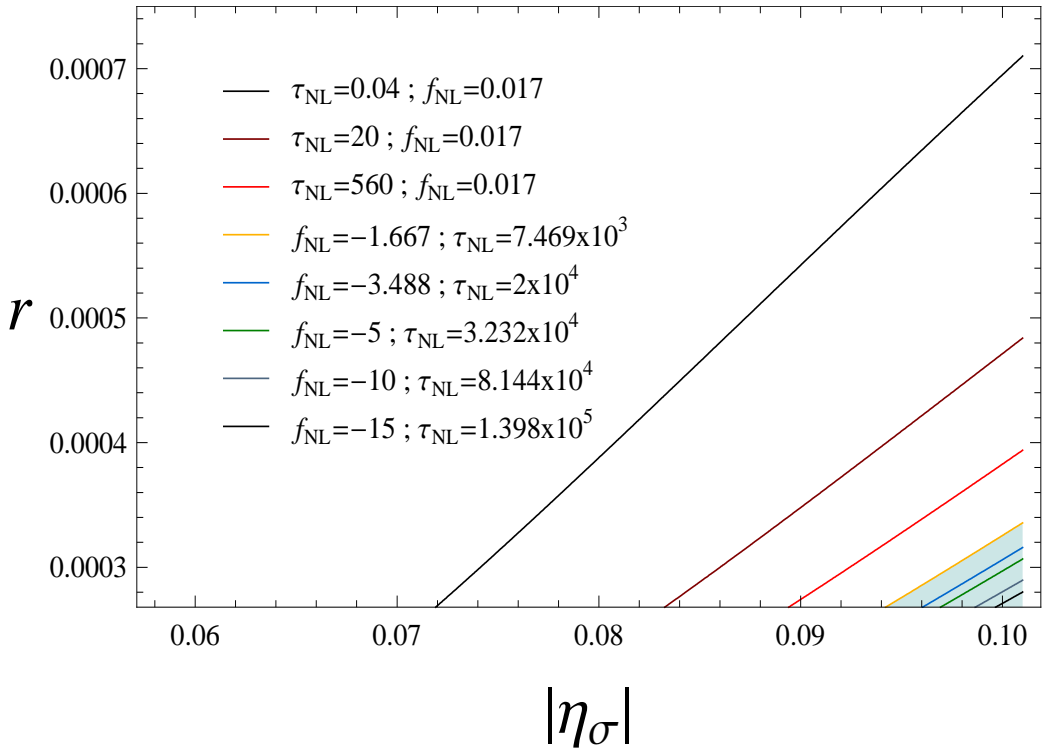}
\end{center}
\caption{Contours of both $f_{NL}$ and $\tau_{NL}$ in the $r$ vs $|\eta_\sigma|$ plot. The intermediate (high) $\phi_\star$ region corresponds to the shaded (white) region. Lines for constant $\tau_{NL}$ almost exactly match lines for constant $f_{NL}$. According to this figure, and to the present observational status, {\it non-gaussianity is more likely to be detected through the trispectrum than through the bispectrum}, for the inflationary model studied in this paper with concave downward potential, and from the WMAP, PLANCK, and even the 21 cm background anisotropies observations. These lines also show some {\it consistency relations between the values of $f_{NL}$ and $\tau_{NL}$} that will be useful at testing the inflationary model considered with concave downward potential against observations.}
\label{fig3}
\end{figure*}

In the companion paper \cite{cogollo}, we studied $f_{NL}$ for the case when $\zeta$ is generated during inflation, $B_\zeta$ is dominated by the one-loop correction, and $P_\zeta$ is dominated by the tree-level term. Fig. \ref{fig2} shows the results found.  The WMAP \cite{komatsu1} (and also PLANCK \cite{komatsu}) observationally allowed $2\sigma$ range of values for negative $f_{NL}$, $-9 < f_{NL}$, is completely inside the intermediate $\phi_\star$ region\footnote{The intermediate $\phi_\star$ $T$-region (where $T_\zeta$ is dominated by the one-loop correction and $P_\zeta$ is dominated by the tree-level term) encloses the intermediate $\phi_\star$ region (where $B_\zeta$ is dominated by the one-loop correction and $P_\zeta$ is dominated by the tree-level term).}. More negative values for $f_{NL}$, up to $f_{NL} = -20.647$, are consistent within our framework for the intermediate $\phi_\star$ region, but they are ruled out from observation. Fig. \ref{fig3} shows both Figs. \ref{fig1} and \ref{fig2} in the same plot. Incidentally, for the available parameter window, lines for constant $\tau_{NL}$ almost exactly match lines for constant $f_{NL}$. Thus, it is possible to see that, according the observational status presented in Subsection 2.2 of Ref. \cite{cogollo}, {\it non-gaussianity is more likely to be detected through the trispectrum than through the bispectrum}, for the inflationary model studied in this paper with concave downward potential, and from the WMAP, PLANCK, and even the 21 cm background anisotropies observations. Fig. \ref{fig3} also shows some {\it consistency relations between the values of $f_{NL}$ and $\tau_{NL}$} that will be useful at testing the inflationary model considered with concave downward potential against observations.  For instance, if WMAP detected non-gaussianity through the trispectrum with $\tau_{NL} \geq 8 \times 10^4$ at the $2\sigma$ level, the slow-roll inflationary model with concave downward potential considered in this paper would be ruled out since the predicted $f_{NL}$ would be outside the current observational interval.

Similar to the $f_{NL}$ case studied in the companion paper \cite{cogollo}, it is interesting to see a slow-roll inflationary model with canonical kinetic terms where large, {\it and observable}, values for $\tau_{NL}$ may be obtained (in contrast to the expected $\tau_{NL} \sim \mathcal{O} (r)$ from the tree-level calculation \cite{seery3,ssv}). So we conclude that {\it if} $T_\zeta$ {\it is dominated by the one-loop correction but} $P_\zeta$ {\it is dominated by the tree-level term, sizeable non-gaussianity is generated even if} $\zeta$ {\it is generated during inflation}.  We also conclude, from looking at the small values that the tensor to scalar ratio $r$ takes in figure \ref{fig3} compared with the present technological bound $r \gsim 10^{-3}$ \cite{friedman}, that {\it for non-gaussianity to be observable in this model, primordial gravitational waves must be undetectable}.

\subsection{$\zeta$ not generated during inflation ($\mathcal{P}_\zeta \ll \mathcal{P}_\zeta^{\rm obs}$)}
We will assume in this subsection that the fields driving inflation have nothing to do with
the generation of $\zeta$, i.e. $\mathcal{P}_\zeta \ll \mathcal{P}_\zeta^{\rm obs}$;
nevertheless, they will generate the primordial non-gaussianity (see, for instance, Refs.
\cite{boubekeur1,suyama,kawasaki1,langlois,hikage,kawasaki2,kawasaki3,kawakami}). To this
end, the post-inflationary evolution, particularly the generation of $\zeta$, will be
assumed not to generate significant levels of non-gaussianity in comparison with those
generated during inflation.

\subsubsection{$\tau_{NL}$: The intermediate $\phi_\star$ $T$-region} \label{iphitregion}
The level of non-gaussianity $\tau_{NL}$ in this case is given by
\bea
\frac{1}{2}\tau_{NL} &=& \frac{T_\zeta^{1-loop}}{8\pi^6\left[\frac{1}{k_2^3 k_4^3 |{\bf k}_3 + {\bf k}_4|^3} + 23 \;\; {\rm permutations}\right] {(\mathcal{P}_\zeta^{\rm obs})^3}} \nonumber\\
&=& \frac{2\eta_{\sigma}^4}{\eta_{\phi}^8\phi_\star^8} \exp[8N(|\eta_\sigma|-|\eta_\phi|)]\left(\frac{H_\star}{2\pi}\right)({\mathcal P}_\zeta^{\rm obs})^{-3} \ln(kL)\nonumber\\
&=&\frac{2\eta_{\sigma}^4}{\eta_{\phi}^8} \exp[8N(|\eta_\sigma|-|\eta_\phi|)]\left(\frac{m_P}{\phi_\star}\right)^8\left(\frac{r}{8}\right)^4{\mathcal P}_\zeta^{\rm obs} \ln(kL) \nonumber\\
&=&\frac{2\eta_{\sigma}^4}{\eta_{\phi}^4}\left(\frac{1}{2\times10^{-2}}\right)^4 \exp(8N|\eta_\sigma|)\left(\frac{r}{8}\right)^4{\mathcal P}_\zeta^{\rm obs} \ln(kL) \nonumber\\
&\simeq&2.60\times10^{16} \ (nr)^4 \,, \label{taunonzeta}
\eea
where in the last line we have defined the parameter $n$ as the ratio between the two $\eta$
parameters: $n = \eta_\sigma / \eta_\phi$. In the last line, we have chosen for simplicity $|\eta_\sigma| = 0.1$ and $N=62$ so that the non-gaussianity is maximized.

The $\zeta$ series convergence constraints in Eqs. (99) and (100) of Ref. \cite{cogollo}, 
and those in Eqs. (\ref{intc}), (\ref{normtd}), and (\ref{amount}) of this paper, with $|\eta_\sigma| = 0.1$ and $N = 62$, lead to the following conditions that reduce the available parameter space:

\begin{itemize}
\item The perturbative regime constraint $|x| \ll 1$:
\begin{equation}
r \lsim 6.51 \times 10^4 \ n \exp\left[-\frac{12.4}{n}\right] \,. 
\end{equation}
\item The perturbative regime constraint $|y| \ll 1$:
\begin{equation}
r \lsim \frac{2.68 \times 10^{-1}}{n^2} \,.  
\end{equation}
\item The $P_\zeta$ dominated by the tree-level term constraint:
\begin{equation}
r \lsim \frac{1.10 \times 10^{-4}}{n} \exp\left[\frac{12.4}{n}\right] \,. 
\end{equation}
\item The $T_\zeta$ dominated by the one-loop correction constraint:
\begin{equation}
r \gsim \frac{4.68 \times 10^{-12}}{n^3} \exp\left[\frac{37.2}{n}\right] \,. \label{plnon1}
\end{equation}
\item The spectrum normalisation constraint:
\begin{equation}
r \lsim \frac{1.6 \times 10^{-4}}{n} \exp\left[-\frac{12.4}{n}\right] \,. \label{sncnonzeta}
\end{equation}
\end{itemize}

\begin{figure*} [t]
\begin{center}
\includegraphics[width=15cm,height=10cm]{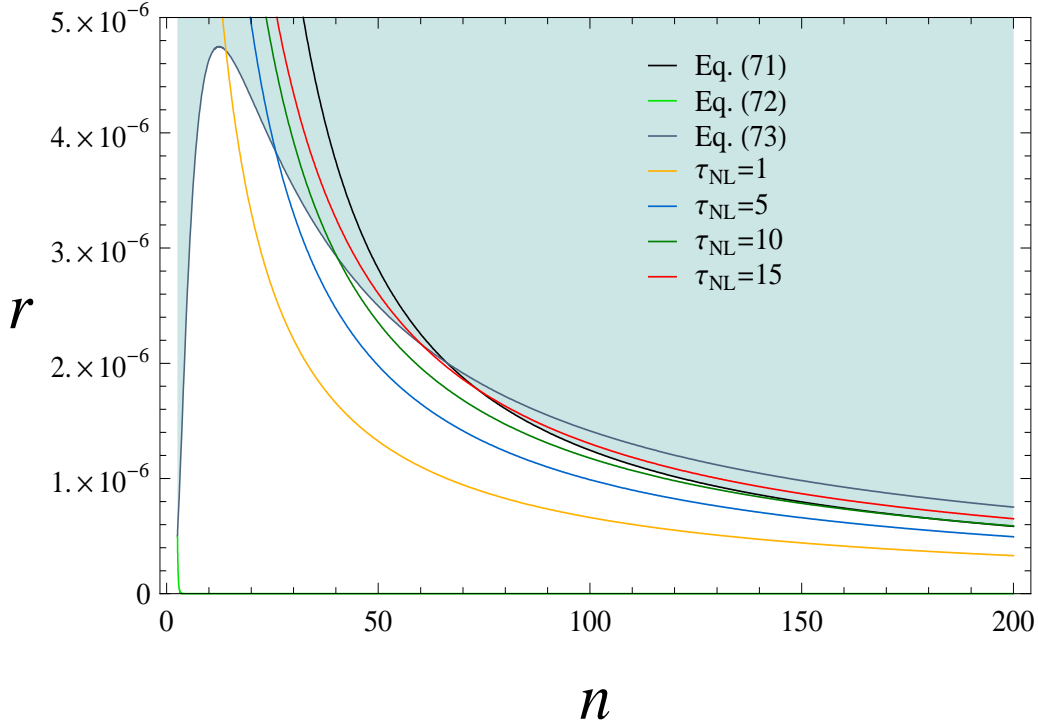}
\end{center}
\caption{Contours of $\tau_{NL}$ in the $r$ vs $n$ plot, for $2.58 \leq n \leq 200$, when $\zeta$ is not generated during inflation. The allowed parameter space corresponds to the white region. The constraint in Eq. (\ref{plnon1}) almost matches (visually) the horizontal axis. The largest possible value $\tau_{NL}$ may take in this range is 15.}
\label{fig4}
\end{figure*}

\begin{figure*} [t]
\begin{center}
\includegraphics[width=15cm,height=10cm]{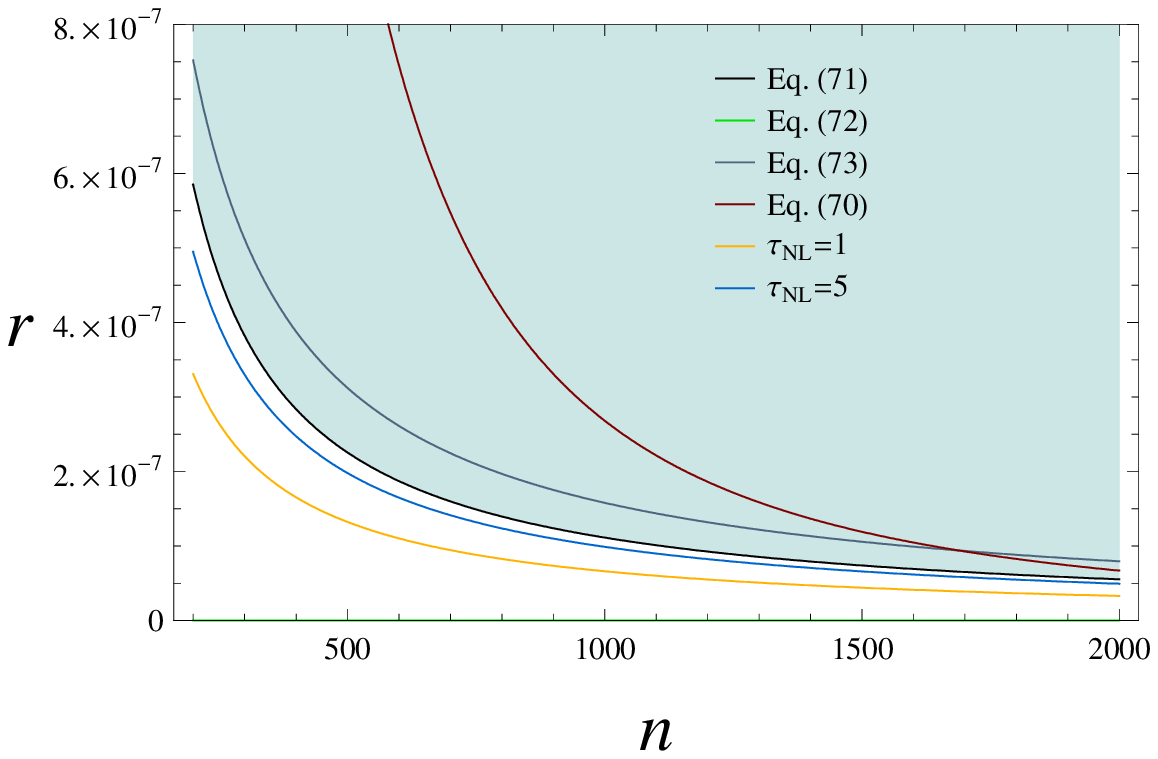}
\end{center}
\caption{Contours of $\tau_{NL}$ in the $r$ vs $n$ plot, for $200 \leq n \leq 2000$, when $\zeta$ is not generated during inflation. The allowed parameter space corresponds to the white region. The constraint in Eq. (\ref{plnon1}) matches (visually) the horizontal axis. The largest possible value $\tau_{NL}$ may take in this range is a bit higher than 5.}
\label{fig5}
\end{figure*}

\begin{figure*} [t]
\begin{center}
\includegraphics[width=15cm,height=10cm]{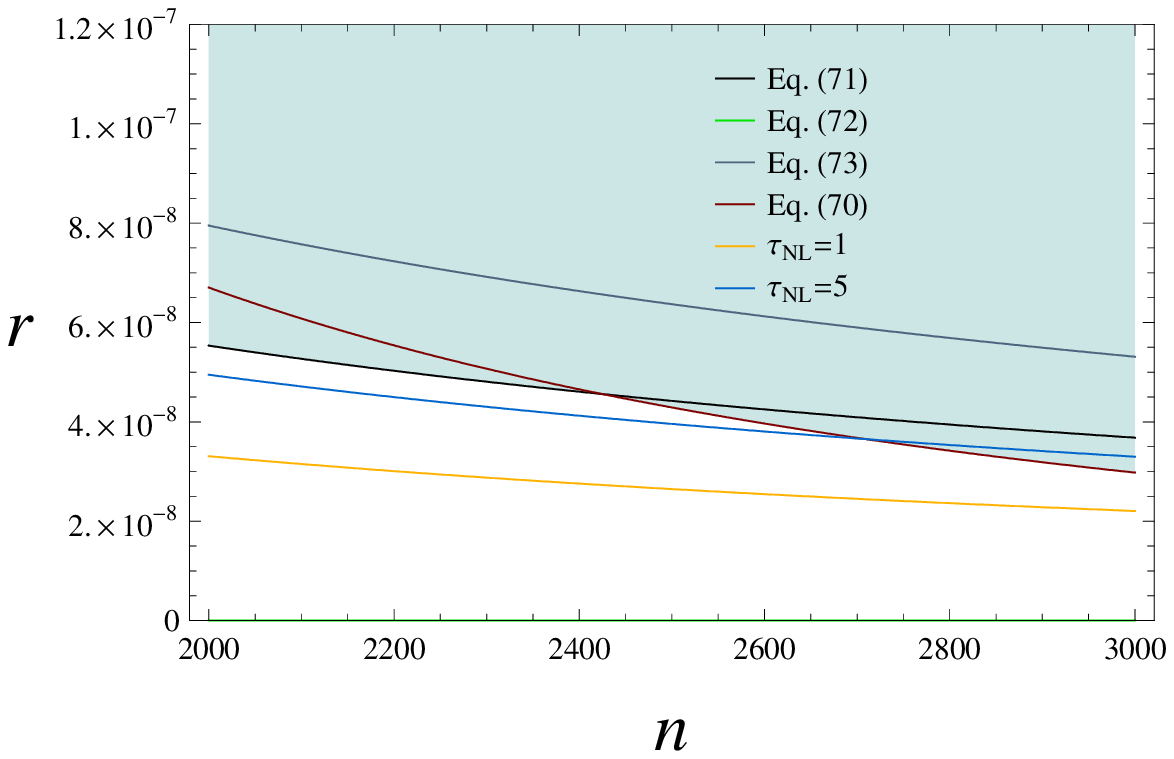}
\end{center}
\caption{Contours of $\tau_{NL}$ in the $r$ vs $n$ plot, for $2000 \leq n \leq 3000$, when $\zeta$ is not generated during inflation. The allowed parameter space corresponds to the white region. The constraint in Eq. (\ref{plnon1}) matches (visually) the horizontal axis. The largest possible value $\tau_{NL}$ may take in this range is a bit higher than 5.}
\label{fig6}
\end{figure*}

Analysing these expressions, we conclude that the constraint in the first item is automatically satisfied once the constraint in the fifth item is satisfied. Moreover, from the constraint in the fifth item, we see that the highest possible value $r$ may take is $4.75 \times 10^{-6}$. And finally, to make the constraint in the fourth item consistent with the constraints in the second, third, and fifth items, the lower bound $n \gsim 2.58$ is required. The resulting available parameter window, together with the lines for constant values of $\tau_{NL}$, $\tau_{NL} = 1,5,10,15$, is presented in Fig. \ref{fig4} for $2.58 \leq n \leq 200$. Fig. \ref{fig5} shows the range $200 \leq n \leq 2000$ with the lines $\tau_{NL} = 1,5$, while Fig. \ref{fig6} shows the range $2000 \leq n \leq 3000$ also with the lines $\tau_{NL} = 1,5$. As the figures reveal, {\it when $T_\zeta$ is dominated by the one-loop correction and $P_\zeta$ is dominated by the tree-level term, large values for $\tau_{NL}$ are obtained although not so large as in the case where $\zeta$ is generated during inflation}.  Indeed, an upper bound on $\tau_{NL}$, according to Eqs. (\ref{taunonzeta}) and (\ref{sncnonzeta}), is $34.078$ when $n \rightarrow \infty$.

We conclude that, even if $\zeta$ is not generated during inflation, we may find {\it observable} values for $\tau_{NL}$. However, such observable values could only be observed by the 21 cm background anisotropies at the $1\sigma$ level according to the observational status presented in Subsection 2.2 of Ref. \cite{cogollo}. We also conclude that, {\it for non-gaussianity to be observable, primordial gravitational waves must be undetectable}.

\subsubsection{$f_{NL}$: the intermediate $\phi_\star$ region} \label{fnlafter}
In the companion paper \cite{cogollo} we only studied the non-gaussianity imprinted in the parameter $f_{NL}$ when $\zeta$ is generated during inflation. To that end, we found in Ref. \cite{cogollo} that the regions of allowed values for $\phi_\star$ also depend on the tree-level or one-loop dominance in $P_\zeta$ and/or $B_\zeta$. We found three different possibilities:

\begin{itemize}
\item The low $\phi_\star$ region, given when both $B_\zeta$ and $P_\zeta$ are dominated by the one-loop corrections, is characterized by (see Eq. (60) of Ref. \cite{cogollo})
\begin{equation}
\left(\frac{\phi_\star}{m_P}\right)^2 \ll \frac{r \mathcal{P}_\zeta^{\rm obs}}{8} \frac{\eta_\sigma^2}{\eta_\phi^2} \exp[4N(\eta_\phi-\eta_\sigma)] \,.
\label{lowphif}
\end{equation}

\item The intermediate $\phi_\star$ region, given when $B_\zeta$ is dominated by the one-loop correction while $P_\zeta$ is dominated by the tree-level term, is characterized by (see Eq. (63) of Ref. \cite{cogollo})
\begin{equation}
\frac{r \mathcal{P}_\zeta^{\rm obs}}{8} \frac{\eta_\sigma^2}{\eta_\phi^2} \exp[4N(\eta_\phi-\eta_\sigma)] \ll \left(\frac{\phi_\star}{m_P}\right)^2 \ll \frac{r \mathcal{P}_\zeta^{\rm obs}}{8} \frac{\eta_\sigma^3}{\eta_\phi^3} \exp[6N(\eta_\phi-\eta_\sigma)] \,. 
\label{intcf}
\end{equation}

\item The high $\phi_\star$ region, given when both $B_\zeta$ and $P_\zeta$ are dominated by the tree-level terms, is characterized by (see Eq. (66) of Ref. \cite{cogollo})
\begin{eqnarray}
\left(\frac{\phi_\star}{m_P}\right)^2 \gg \frac{r \mathcal{P}_\zeta^{\rm obs}}{8} \frac{\eta_\sigma^3}{\eta_\phi^3} \exp[6N(\eta_\phi-\eta_\sigma)] \,.
\label{highphif}
\end{eqnarray}
\end{itemize}


The low and high $\phi_\star$ regions will be studied in Appendix \ref{app}.
Regarding the intermediate $\phi_\star$ region, the level of non-gaussianity is in this case given by
\bea
\frac{6}{5} f_{NL} &=& \frac{B_\zeta^{1-loop}}{4\pi^4 \frac{\sum_i k_i^3}{\prod_i k_i^3} (\mathcal{P}_\zeta^{\rm obs})^2} = \frac{\eta_\sigma^3}{\eta_\phi^6 \phi_\star^6} \exp[6N(|\eta_\sigma|-|\eta_\phi|)] \left(\frac{H_\star}{2\pi}\right)^6 (\mathcal{P}_\zeta^{\rm obs})^{-2} \ln(kL) \nonumber \\
&=& \frac{\eta_\sigma^3}{\eta_\phi^6} \exp[6N(|\eta_\sigma|-|\eta_\phi|)] \left(\frac{m_P}{\phi_\star}\right)^6 \left(\frac{r}{8}\right)^3 \mathcal{P}_\zeta^{\rm obs} \ln(kL) \nonumber \\
&=& -\frac{\eta_\sigma^3}{\eta_\phi^3} \left(\frac{1}{2\times10^{-2}}\right)^3 \exp(6N |\eta_\sigma|) \left(\frac{r}{8}\right)^3 \mathcal{P}_\zeta^{\rm obs} \ln(kL) \nonumber \\
&\approx& -8.59 \times 10^{9} (nr)^3 \,, \label{fnonzeta}
\eea
where
in the last line we have chosen again for simplicity $|\eta_\sigma| = 0.1$ and $N = 62$ so that the non-gaussianity is maximized.

Since the spectrum normalisation constraint in Eq. (\ref{sncnonzeta}) equally applies to this case, we conclude from it and from Eq. (\ref{fnonzeta}) that an upper bound on $|f_{NL}|$ is $2.93 \times 10^{-2}$ when $n \rightarrow \infty$. $f_{NL}$ is, of course, unobservable.  We conclude that {\it when $\zeta$ is not generated during inflation, but the primordial non-gaussianity is, it is impossible to detect non-gaussianity through the bispectrum}. However, in view of Subsubsection \ref{iphitregion}, {\it it is possible to detect it through the trispectrum}. 

\section{Conclusions} \label{concl}

Is it reasonable to study the primordial curvature perturbation $\zeta$ by identifying it
with a truncated $\delta N$ series expansion? Is it actually possible to cut out with
confidence such a series at some specific order? Is it true that all of the slow-roll
inflationary models with canonical kinetic terms produce primordial non-gaussianity
supressed by the slow-roll parameters? Are the loop corrections in cosmological perturbation
theory always smaller than the tree-level terms?  We have addressed these questions in this
paper and its companion \cite{cogollo}, answering all of them by paying particular attention
to a special slow-roll inflationary model. 
The $\zeta$ series expansion is indeed a powerful tool to study the statistical
descriptors of $\zeta$; nevertheless, we should seek the convergence radius in order
not to obtain results that actually have nothing to do with $\zeta$. We may cut out the
series but, to be completely sure about the precision of our approximations, we have to
study the conditions for the existence of a perturbative regime. Non-gaussianity in
slow-roll inflationary models with canonical kinetic terms is not always suppressed by the
slow-roll parameters; we have seen this at tree level for $f_{NL}$ in
Refs. \cite{alabidi1,byrnes3}, and considering loop corrections for $f_{NL}$ in
Refs. \cite{cogollo,bch2} and for $\tau_{NL}$ in Ref. \cite{bch2} and the present paper.
The statement presented in Ref. \cite{bch2} about the suppression of the loop corrections
against the tree-level terms when considering classicality was analyzed in this paper and
argued to be too strongly stated leading to nongeneral conclusions. The probability that a
typical observer sees a non-gaussian distribution in the model considered in this paper and
its companion \cite{cogollo} was investigated and found to be non-negligible. Finally, as far
as we have investigated, the loop corrections in cosmological perturbation theory are not
always smaller than the tree-level terms; in fact, when they become the leading contributions,
a surprising phenomenology appears in front of our eyes.

\subsection*{Acknowledgments}
This work is supported by COLCIENCIAS grant No. 1102-333-18674 CT-174-2006, DIEF (UIS) grant
No. 5134, and by the ECOS-NORD programme grant No. C07P02. Y.R. acknowledges Misao Sasaki,
David H. Lyth, Christian Byrnes, Neil Barnaby, Ki-Young Choi, Filippo Vernizzi, Eiichiro
Komatsu, Fuminobu Takahashi, Kazunori Kohri, and Chia-Min Lin for useful conversations.
We acknowledge the invaluable help and suggestions of two anonymous referees. 
Y.R. also acknowledges the hospitality of the IPMU - Institute for the Physics and Mathematics
of the Universe at the University of Tokyo (Japan) during the workshop ``{\it Focus Week on
Non-Gaussianities in the Sky}'' (April 2009), and the YITP - Yukawa Institute for Theoretical
Physics at Kyoto University (Japan) during the GCOE/YITP workshop YITP-W-09-01 on
``{\it Non-Linear Cosmological Perturbations}'' (April 2009), where the key ideas for the
revised version of this paper were unveiled. C.A.V.-T. acknowledges the Department of Physics
at Lancaster University (UK) where part of this work was done.


\appendix

\section{Complementary cases} \label{app}

In this appendix, we will show complementary cases to the ones studied in Sections \ref{constraints}
and \ref{endcal}, namely, 1. $\tau_{NL}$ when $\zeta$ is generated during inflation
and both $P_\zeta$ and $T_\zeta$ are dominated by the tree-level terms, 2. $f_{NL}$ and
$\tau_{NL}$ when $\zeta$ is not generated during inflation and both $P_\zeta$, $B_\zeta$, and
$T_\zeta$ are dominated by the one-loop corrections, 3. $\tau_{NL}$ when $\zeta$ is not
generated during inflation and both $P_\zeta$ and $T_\zeta$ are dominated by the tree-level
terms, and 4. $f_{NL}$ when $\zeta$ is not generated during inflation and both $P_\zeta$ and
$B_\zeta$ are dominated by the tree-level terms.

\subsection{Tree-level or loop dominance}

\subsubsection{Both $T_\zeta$ and $\mathcal{P}_\zeta$ are dominated by the one-loop corrections}
Comparing Eqs. (\ref{pt}) with (\ref{pl}) and Eqs. (\ref{tt}) with (\ref{tl}) we require in this case that
\bea
\frac{\eta_\sigma^2}{\eta_\phi^2} \exp[4N(\eta_\phi-\eta_\sigma)] &\gg& \frac{1}{\frac{1}{\phi_\star^2} \left(\frac{H_\star}{2\pi}\right)^2} \,, \\
4\frac{\eta_\sigma^4}{\eta_\phi^4} \exp[8N(\eta_\phi-\eta_\sigma)] &\gg& \frac{1}{\frac{1}{\phi_\star^2} \left(\frac{H_\star}{2\pi}\right)^2} \,,
\eea
in which case only the first inequality is required. Employing the definition for the tensor
to scalar ratio $r$ introduced in Eq. (\ref{rts}), we can write such inequality as
\be
\left(\frac{\phi_\star}{m_P}\right)^2 \ll \frac{r \mathcal{P}_\zeta^{\rm obs}}{8} \frac{\eta_\sigma^2}{\eta_\phi^2} \exp[4N(\eta_\phi-\eta_\sigma)] \,.
\label{ptloop}
\ee
From now on we will name the parameter window described by Eq. (\ref{ptloop}) as the low
$\phi_\star$ $T$-region, since the latter represents a region of allowed values for
$\phi_\star$ limited by an upper bound.

\subsubsection{Both $T_\zeta$ and $\mathcal{P}_\zeta$ are dominated by the tree-level terms} \label{hiphit}
Comparing Eqs. (\ref{pt}) with (\ref{pl}) and Eqs. (\ref{tt}) with (\ref{tl}) we require in this case that
\bea
\frac{\eta_\sigma^2}{\eta_\phi^2} \exp[4N(\eta_\phi-\eta_\sigma)] &\ll& \frac{1}{\frac{1}{\phi_\star^2} \left(\frac{H_\star}{2\pi}\right)^2} \,, \\
4\frac{\eta_\sigma^4}{\eta_\phi^4} \exp[8N(\eta_\phi-\eta_\sigma)] &\ll& \frac{1}{\frac{1}{\phi_\star^2} \left(\frac{H_\star}{2\pi}\right)^2} \,,
\eea
in which case only the second inequality is required.  Employing the definition for the tensor to scalar ratio $r$ introduced in Eq. (\ref{rts}), we can write such an inequality as
\bea
\left(\frac{\phi_\star}{m_P}\right)^2 \gg \frac{r \mathcal{P}_\zeta^{\rm obs}}{8} \frac{4\eta_\sigma^4}{\eta_\phi^4} \exp[8N(\eta_\phi-\eta_\sigma)] \,.
\label{highphi}
\eea
From now on, we will name the parameter window described by Eq. (\ref{highphi}) as the high $\phi_\star$ $T$-region, since the latter represents a region of allowed values for $\phi_\star$ limited by a lower bound.

\subsection{The normalisation of the spectrum}

\subsubsection{$\zeta$ not generated during inflation ($\mathcal{P}_\zeta \ll \mathcal{P}_\zeta^{\rm obs}$) and $\mathcal{P}_\zeta$ dominated by the one-loop correction}
According to Eqs. (\ref{pl}) and (\ref{rts}) we have in this case
\bea
\mathcal{P}_\zeta^{1-loop} &=& \frac{\eta_\sigma^2}{\eta_\phi^4 \phi_\star^4} \exp[4N(\eta_\phi - \eta_\sigma)] \left(\frac{H_\star}{2\pi}\right)^4 \ln(kL) \nonumber \\
&=& \frac{\eta_\sigma^2}{\eta_\phi^4} \exp[4N(\eta_\phi - \eta_\sigma)] \left(\frac{m_P}{\phi_\ast}\right)^4 \left(\frac{r \mathcal{P}_\zeta^{\rm obs}}{8}\right)^2 \ln(kL) \,,
\eea
which reduces to
\be
\left(\frac{\phi_\star}{m_P}\right)^4 \gg \left(\frac{r}{8}\right)^2 \mathcal{P}_\zeta^{\rm obs} \frac{\eta_\sigma^2}{\eta_\phi^4} \exp[4N(\eta_\phi - \eta_\sigma)] \ln(kL)\,, \label{norml}
\ee
where $\mathcal{P}_\zeta^{\rm obs}$ must be replaced by the observed value $(\mathcal{P}_\zeta^{\rm obs})^{1/2} = (4.957 \pm 0.094) \times 10^{-5}$ \cite{bunn}.


\subsection{$\zeta$ generated during inflation ($\mathcal{P}_\zeta = \mathcal{P}_\zeta^{\rm obs}$)}

\subsubsection{The high $\phi_\star$ $T$-region}
According to the expressions in Eqs. (\ref{pt}) and (\ref{tt}) of this paper, and Eqs. (18), (20), and (78) of Ref. \cite{cogollo}, the value of $\tau_{NL}$ is in this case
\be
\frac{1}{2}\tau_{NL} = \frac{T^{tree}_\zeta}{8\pi^6 \left[\frac{1}{k_2^3 k_4^3 |{\bf k}_3 + {\bf k}_4|^3} + 23 \;\; {\rm permutations}\right] (\mathcal{P}^{tree}_\zeta)^3}=\frac{1}{2}\eta_\phi^2 = 2\times10^{-4}\,,
\ee
in agreement with the general expectations of Ref. \cite{byrnes2} for slow-roll inflationary models with canonical kinetic terms where only the tree-level contributions are considered and the field perturbations are assumed to be gaussian. This result is of no observational interest because the generated non-gaussianity is too small to be observable.

\subsection{$\zeta$ not generated during inflation ($\mathcal{P}_\zeta \ll \mathcal{P}_\zeta^{\rm obs}$)}

\subsubsection{$\tau_{NL}$: The low $\phi_\star$ $T$-region}\label{nolow}
It is possible, in principle, that $P_\zeta$ is dominated by the one-loop correction as long as $\zeta$ is not generated during inflation.  Thus, the observed spectral index constraint is no longer required and, therefore, the low $\phi_\star$ $T$-region is in principle viable.

Combining the conditions in Eqs. (\ref{ptloop}) and (\ref{norml}) with the expression for the number of e-folds in Eq.
(\ref{amount}), we get
\be
1\lsim \frac{rn^2}{16} {\mathcal P}_\zeta^{\rm obs} \exp[N|\eta_\sigma|(4-2/n)]\,,\label{no1}
\ee
and
\be
1\gsim10^6\left(\frac{rn}{16}\right)^2{\mathcal P}_\zeta^{\rm obs}\exp(4N|\eta_\sigma|)\,,\label{no2}
\ee
where we have defined the parameter $n$ as in Subsubsection \ref{iphitregion}. These two expressions lead to 
\be
rn\lsim1.6\times10^{-5}|\eta_\sigma|\exp(-2N|\eta_\sigma|/n) \,,
\ee
as a necessary but not sufficient condition to satisfy both Eqs. (\ref{no1}) and (\ref{no2}). However, by introducing such a condition in Eq. (\ref{no1}), we see that the latter translate into the following constraint:
\be
1\lsim 10^{-6}|\eta_\sigma|^2{\mathcal P}_\zeta^{\rm obs}\exp[4N|\eta_\sigma|(1-1/n)] \,.
\ee
The previous expression is impossible to satisfy because the highest value  the right hand side may take is for $n\rightarrow \infty$ and, of course, $\eta_\sigma=0.1$ and $N=62$. Such a value, $1.45\times10^{-6}$, is much less than one. We conclude that this case is of no interest, because it is impossible to satisfy the normalisation spectrum condition in Eq. (\ref{norml}).

\subsubsection{$f_{NL}$: The low $\phi_\star$ region}
From Eqs. (\ref{ptloop}) and (\ref{lowphif}) we see that the low $\phi_\star$ region and the
low $\phi_\star$ $T$-region are exactly the same, being only constrained by the fact that
$P_\zeta$ is dominated by the one-loop correction.  Thus, the obtained conclusions in
Subsubsection \ref{nolow} equally apply. Therefore, this case is of no interest, because it
is impossible to satisfy the normalisation spectrum condition in Eq. (\ref{norml}) for the
low $\phi_\star$ region. 

\subsubsection{$\tau_{NL}$: The high $\phi_\star$ $T$-region}
This case is of no interest because the generated non-gaussianity is too small to be observable:
\bea
\frac{1}{2}\tau_{NL} &=& \frac{T_\zeta^{tree}}{8\pi^6\left[\frac{1}{k_2^3 k_4^3 |{\bf k}_3 + {\bf k}_4|^3} + 23 \;\; {\rm permutations}\right] {(\mathcal{P}_\zeta^{\rm obs})^3}} \nonumber\\
&=&  \frac{T_\zeta^{tree}}{8\pi^6\left[\frac{1}{k_2^3 k_4^3 |{\bf k}_3 + {\bf k}_4|^3} + 23 \;\; {\rm permutations}\right] \left(\mathcal{P}_\zeta^{tree}\right)^3} \frac{\left(\mathcal{P}_\zeta^{tree}\right)^3}{(\mathcal{P}_\zeta^{\rm obs})^3}\;=\; \frac{1}{2} |\eta_\phi|^2 \frac{\left(\mathcal{P}_\zeta^{tree}\right)^3}{(\mathcal{P}_\zeta^{\rm obs})^3}\nonumber\\
&\Rightarrow& \tau_{NL}\;\ll\;|\eta_\phi|^2 \,,
\eea
according to the expressions in Eqs. (\ref{pt}) and (\ref{tt}).

\subsubsection{$f_{NL}$: The high $\phi_\star$ region}
This case is of no interest because the generated non-gaussianity is too small to be observable:
\be
\frac{6}{5} f_{NL} = \frac{B_\zeta^{tree}}{4\pi^4 \frac{\sum_i k_i^3}{\prod_i k_i^3} (\mathcal{P}_\zeta^{\rm obs})^2} =  \frac{B_\zeta^{tree}}{4\pi^4 \frac{\sum_i k_i^3}{\prod_i k_i^3} \left(\mathcal{P}_\zeta^{tree}\right)^2} \frac{\left(\mathcal{P}_\zeta^{tree}\right)^2}{(\mathcal{P}_\zeta^{\rm obs})^2} = |\eta_\phi| \frac{\left(\mathcal{P}_\zeta^{tree}\right)^2}{(\mathcal{P}_\zeta^{\rm obs})^2} \ll |\eta_\phi| \,,
\ee
according to the expressions in Eqs. (\ref{pt}) and (\ref{bt}).

\renewcommand{\refname}{{\large References}}

\end{document}